\documentclass[conference]{IEEEtran}
\usepackage{times}
\usepackage{balance} 
\usepackage{relsize}
\usepackage{scalerel}
\usepackage{float}
\usepackage[ruled]{algorithm2e}
\usepackage{silence}
\usepackage{amsmath}
\usepackage{xcolor}
\usepackage{booktabs}
\usepackage[numbers]{natbib}
\usepackage{empheq}

\usepackage{multirow}
\usepackage[bookmarks=true]{hyperref}

\newcommand{\BibTeX}{\rm B\kern-.05em{\sc i\kern-.025em b}\kern-.08em\TeX}
\newcommand{\rotate}[1]{\rotatebox[origin=c]{90}{\parbox[c]{1cm}{\centering #1}}}

\newcommand{\blue}[1]{\color{blue}#1\normalcolor}
\newcommand{\red}[1]{\color{red}#1\normalcolor}

\newcommand{\lblue}[1]{\color{black}#1\normalcolor}
\newcommand{\lred}[1]{\color{black}#1\normalcolor}

\DeclareMathOperator*{\argmin}{argmin}

\begin{document}

\title{Online Competitive Information Gathering for Partially Observable Trajectory Games}



%
\author{\authorblockN{Mel Krusniak\authorrefmark{1},
{Hang Xu}\authorrefmark{1},
{Parker Palermo}\authorrefmark{1}, 
and {Forrest Laine}\authorrefmark{1}}
\authorblockA{\authorrefmark{1} Vanderbilt University, Nashville, TN 37212 \\ Email: mel.krusniak@vanderbilt.edu}
}





\maketitle
\newcommand\mel[1]{\textcolor{purple}{Mel: #1}}

\begin{abstract}

Game-theoretic agents must make plans that optimally gather information about their opponents. These problems are modeled by partially observable stochastic games (POSGs), but planning in fully continuous POSGs is intractable without heavy offline computation or assumptions on the order of belief maintained by each player. We formulate a finite history/horizon refinement of POSGs which admits competitive information gathering behavior in trajectory space, and through a series of approximations, we present an online method for computing rational trajectory plans in these games which leverages particle-based estimations of the joint state space and performs stochastic gradient play. We also provide the necessary adjustments required to deploy this method on individual agents. The method is tested in continuous pursuit-evasion and warehouse-pickup scenarios (alongside extensions to $N>2$ players and to more complex environments with visual and physical obstacles), demonstrating evidence of active information gathering and outperforming passive competitors.
\end{abstract}

\IEEEpeerreviewmaketitle


\section{Introduction}

In this work, we consider rational online trajectory planning in non-cooperative, imperfect-information settings. While approximate solutions can be found to some partially observable multi-robot problems through e.g. reinforcement learning, modeling the information interactions of noncooperative robots in realistic spaces during execution remains difficult. For instance, robots in adversarial scenarios can leverage the sensing limitations of their adversaries to mislead or misdirect. These interactions are critical to intelligent behavior: intelligent agents seek out and use information, and anticipate their counterparts doing the same.

Observation interactions such as this are often modeled with either \textit{extensive form games} or \textit{partially observable stochastic games}, two closely related formulations \cite{yang2020overview}. These game-theoretic formulations allow for provably correct behavior with respect to adversarial information gathering and deception. Using these formulations to solve discrete parlor games like poker \cite{tammelin2015solving} is well studied, but at present, it remains an open challenge to do so live in continuous spaces as required in robotics without extensive offline computation and training. Our work addresses this case, presenting a method to plan information-aware strategies in the manner of model predictive game play (or MPGP --- an extension of model predictive control to game-theoretic multi-robot environments) \cite{cleac2019algames}. We contend that this work is complementary to and compatible with offline methods \cite{becker_bridging_2024}, and even when online methods are resource-intensive, they allow us to consider what is required to compute solutions from scratch. Our method is not real-time, though we demonstrate that real-time performance is within reach.

\begin{figure}
    \centering
    \includegraphics[width=\linewidth,trim={0 0 0 0},clip]{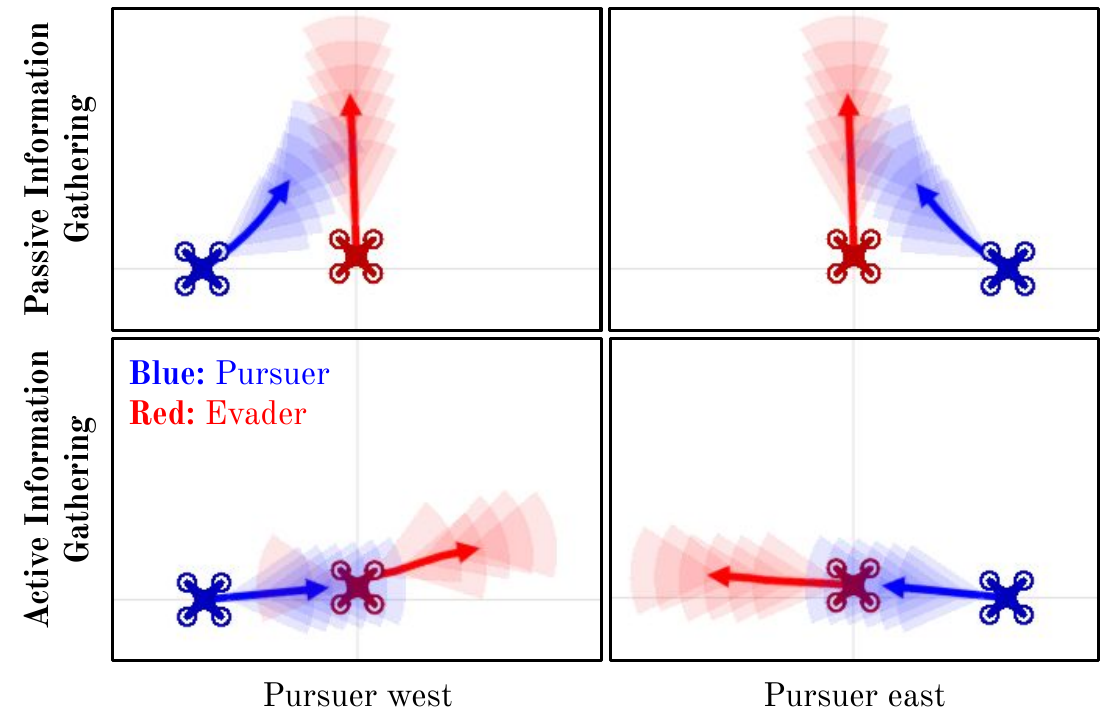}
    \caption{Illustration of information gathering modalities in a pursuit-evasion game ``\textsc{Tag}''. Cones indicate field of view at each planned step. The \blue{pursuer}'s initial position (``east'' / ``west'') is random and unknown to the \red{evader}.} 
    \label{fig:passive_vs_active_gathering}
\end{figure}

As a concrete example, consider a pursuit-evasion scenario between two free-moving UAVs with mounted cameras. Each UAV observes its opponent's location if its opponent is within its field of view (assumed to face the direction of motion), and observes random noise otherwise. The pursuer starts with equal probability in one of two locations  --- unknown to the evader, which starts between them. In an ideal, information-aware trajectory plan, the evader briefly turns to check one location, then executes the remainder of the plan based on its observation: a desirable behavior known as \textit{active information gathering}. (In the alternative, \textit{passive information gathering}, plans are generated without considering potential future observations.) Figure \ref{fig:passive_vs_active_gathering} visualizes the distinction in the UAV pursuit-evasion scenario, where each column shows a possible ground truth pursuer location. 

Standard MPGP considers only perfect information games, so this work develops a variant which uses imperfect information games to permit active information gathering. We present a simple, live planner for these scenarios which permits competition through information as well as in trajectory space using a particle representation of the joint distribution of observation histories and states.

\subsection{Technical Contributions}

This work presents the following technical contributions:

\begin{itemize}
    \item A novel enhancement of model-predictive game play to handle imperfect information games, addressing POSGs over a finite history and horizon;
    \item Description of a simple online solver method for this setup, with additions to permit execution in more realistic robotic planning contexts;
    \item Evidence that this method achieves active information gathering in pursuit-evasion and warehouse-pickup robotics scenarios, alongside extensions to $N$-player games and more realistic setups with visual and physical obstacles; and
    \item Experimental data regarding timing and equilibrium agreement using our method, as necessary considerations for practical game-theoretic multi-agent robotics.  
\end{itemize}


\section{Background}

Our work is influenced by several overlapping formulations for imperfect information rational planning. 
\begin{itemize}
    \item \textit{Partially observable stochastic games} (POSGs) are used in modern training of competitive agents. 
    \item \textit{Extensive form games} (EFGs), closely related to POSGs, are foundational for no-regret algorithms regarding agent interactions but are difficult to apply to large problems. 
    \item \textit{Multi-agent belief space planning} adapts belief-based control theory methods to $N$-player environments.
    \item \textit{Interactive POMDPs} are applicable to some specific multi-agent cases with limitations on information modeling.
\end{itemize}

In this section, we address these major influences.

\subsection{POSGs and EFGs}
\textit{Partially observable stochastic games} are $N$-player extensions of POMDPs \cite{albrecht2024_marl_book}, \cite{yang2020overview}. As an extension of planning in POMDPs, planning in POSGs is NP-hard \cite{vlassis2012_pomdps_planning_nphard}, and most POSGs cannot be tractably solved. Nevertheless, modern advances in multi-agent reinforcement learning (MARL) have yielded approaches to finding strong strategies in POSGs, particularly in imperfect information parlor/video games like Stratego \cite{perolat2022mastering}, Starcraft \cite{vinyals2019grandmaster}, and Diplomacy \cite{bakhtin2022mastering}. POSGs are closely related to \textit{extensive form games} (EFGs), which provide a treelike interpretation of planning under imperfect information. Both model similar problems; the exact distinction made varies  \cite{kovavrik2022rethinking} \cite{becker_bridging_2024}. 

Many important multi-agent RL-based methods are informed by POSGs and/or EFGs, including neural fictitious self-play \cite{heinrich2016deep}, deep counterfactual regret minimization \cite{brown2019deep}, and variations on policy space response oracles \cite{lanctot2017unified} \cite{mcaleer_xdo_2021}. In these methods, deep neural networks extend from tabular to continuous contexts and handle large state and action spaces at the cost of expensive offline training. Rather than solving such games in their entirety in this manner, we take an online view, seeking rational behavior in settings where pre-computation is not an option.

\subsection{Belief Space Planning and Model Predictive Game Play}
Given the context of online planning, we also consider control-theoretic work. For instance, belief space planning is an important concept in single-player planning which creates plans in state distributions (belief space) rather than in state space \cite{platt2010belief}. These methods are online, using minimal if any deep learning. Unfortunately, while formulating POMDPs into belief MDPs for this purpose is common, there is no widespread corresponding concept of a ``belief POSG.'' Attributing beliefs to other players (as in the theory of mind \cite{scassellati2002theory}) forms a belief hierarchy: players have beliefs over the state, beliefs over other players' beliefs, beliefs over those beliefs, and so on. Working around this issue imposes limitations. Some methods (e.g., Laine et al. \cite{laine2021multi}) attribute immediate intentions in a planning framework but not formal beliefs. Schwarting et al. \cite{schwarting2021stochastic} presents an LQR-style controller for multiplayer belief-space planning but only considers first-order beliefs, which limits the potential competitive behavior in the information space.

Our work relies on a paradigm called \textit{model predictive game play} (sometimes \textit{game planning}), or MPGP \cite{cleac2019algames} \cite{wang2019game}. Rather than solving a trajectory optimization problem for each horizon as in typical model predictive control, in MPGP, players solve a trajectory equilibrium problem (that is, a game). Existing MPGP approaches are not designed for imperfect information games, though progress has been made by explicitly rewarding information gathering --- for instance, Sadigh et al. \cite{sadigh2016information} do so by rewarding observations that inform a model of human behavior. (Here, we reward agents for seeking information in terms of improvement on the underlying non-informational task rather than introducing a separate reward.)

\subsection{Alternative Multi-Agent POMDPs}

In competitive settings, \textit{interactive POMDPs} (I-POMDPs) \cite{gmytrasiewicz2004interactive} model uncertainty interactions with arbitrarily hierarchical beliefs. This yields nuanced strategies but causes a large belief space that is computationally difficult to handle. Methods to address this include particle-filter approaches (somewhat akin to the approach used in our work) and Monte Carlo tree search variants \cite{han2018learning} \cite{schwartz2022online}. However, these methods were tested only on simple, discrete-space games. Furthermore, I-POMDP-based approaches are fundamentally limited by reasoning at a finite order. Higher-order reasoning (e.g., reasoning about other players' beliefs about one's own belief) is made intractable by an exponential dependency of belief space on belief order. 

\subsection{Summary of Background}
Existing learning-based methods on frameworks like POSGs and EFGs address partially observable settings, but do so via offline training (and sometimes only in discrete spaces). More specific formulations restrict competition over information. Meanwhile, belief-space planning and model-predictive game play address online planning, but are either partially observable or multiplayer; never both. We situate our work between these to consider online game-theoretic planning under imperfect information in full.

In the remainder of this work we use POSGs as our descriptive framework, which suitably express imperfect information while permitting intuitive model-predictive game play.

\section{Formulation}

\newcommand{\tfuture}[0]{{T_\textrm{future}}}
\newcommand{\tpast}[0]{{T_\textrm{past}}}

\subsection{Framework} 
\label{sec:posgs_framework}
Consider a POSG $\mathcal{G}$, as defined as a tuple with the following elements:

\begin{itemize}
    \item $\mathcal{N} := 1..N$, the set of players;
    \item $\mathcal{X}^{(i)}$, the set of states for player $i$ (s.t. $x^{(i)}_t \in \mathcal{X}^{(i)})$;
    \item $\mathcal{A}^{(i)}$, the set of actions for player $i$ (s.t. $a_t^{(i)} \in \mathcal{A}^{(i)})$;
    \item $\mathcal{Z}^{(i)}$, the set of observations for player $i$ (s.t. $z_t^{(i)} \in \mathcal{Z}^{(i)})$;
    \item $T^{(i)}(x^{(i)}_t | x_{t-1}, a_{t-1})$, the state transition for player $i$; 
    \item $O^{(i)}(z^{(i)}_t | x_t)$, the observation model for player $i$; 
    \item $r^{(i)}(x_t)$, the (simplified) reward function for player $i$; and
    \item $X_0(x_0)$, the joint prior distribution over players' states. (This distributes over initial configurations of the game for all players. These initial states may be correlated between players.) 
\end{itemize}

For notational simplicity we assume $x$ is factored by player, though this need not be the case. We index over both players and time: $x^{(i)}$ denotes player $i$'s states for all timesteps, $x_t$ denotes \textit{all} players' states at timestep $t$, and $x$ denotes the joint state at all timesteps (and likewise for actions $a$ and observations $z$). To maintain parity with extensive form games, we speak in terms of cumulative costs, and define them based on instant rewards as $c^{(i)}(x) := -\sum_{t=1}^\infty r^{(i)}(x_t)$.

We make the following high-level assumptions on all POSGs that we consider:

\textbf{Assumption 1 (\textit{Complete POSG and common knowledge}).} $\mathcal{G}$ is fully defined; we do not consider incomplete games or model-free methods. All costs and transitions are public.

\textbf{Assumption 2 (\textit{Differentiability}).} $T^{(i)}$, $O^{(i)}$, and $r^{(i)}$ are differentiable in all variables. This is required to use gradient play to seek equilibria at the planning level.

\textbf{Assumption 3 (\textit{Deterministic Strategies}).} While mixed strategies can elicit competitive improvement in the types of games we consider \cite{peters2022learning}, this improvement is often minor. For simplicity, we only consider pure strategies in this work.

\subsection{Imperfect Information MPGP}
We encounter two major necessities when adapting POSGs to MPGP:
\begin{enumerate}
    \item \textbf{Active information gathering}: Rather than planning single trajectories, players find maps from possible future observation histories to trajectories. (This follows from the concept of information sets in extensive form games; observation histories in POSGs uniquely and perfectly index information sets. It also corresponds to a shift from open- to closed-loop Nash equilibria \cite{bacsar2018handbook}.) 
    \item \textbf{Belief modeling}: Players must model account for the past and future observations of \textit{other} players, and plan accordingly, without invoking the belief hierarchy.
\end{enumerate}

To address the former, we consider finite past and future of lengths $\tpast$ and $\tfuture$ respectively, and optimize mappings from observation histories $z^{(i)}_{[t]} := [z^{(i)}_{t-\tpast+1}, ..., z^{(i)}_t]$ to actions $a^{(i)}_t$. (We represent these mappings as policies $\pi$ with parameters $\theta$.) To address the latter, we require players to track the (fully unconditioned) joint distribution of observation histories and states $q_{\bar{t}}(x_{\bar{t}}, z_{[\bar{t}]})$) from $X_0$ to the planning time $\bar{t}$, a move which we motivate in the section that follows.

The resulting game is
\begin{equation}
    \left \{
        \argmin_{\theta^{(i)}}  
        \int_{\substack{
            {x} \in \mathcal{X}^{1+\tfuture} \\
            {z} \in \mathcal{Z}^{\tpast+1+\tfuture}
        }}
        c^{(i)}(x)
        p(x, z | x_{\bar{t}}, z_{[\bar{t}]})
        q_{\bar{t}}(x_{\bar{t}}, z_{[\bar{t}]})
    \right \}_{i}
    \label{eqn:succinct_tg}
\end{equation}
where
\begin{equation}
    p(x, z | x_{\bar{t}}, z_{[\bar{t}]}) = \prod_{t=\bar{t}}^{\bar{t} + \tfuture} T(x_{t+1} | x_t, \pi_{\theta}(z_{[t]}))\; O(z_t|x_t)
\end{equation}
and $i \in 1..N$. 
\textbf{This game is the core of our approach}; each player $k \in 1..N$ will simultaneously solve all $N$ of the optimization problems in (\ref{eqn:succinct_tg}) once per timestep to generate a plan (and generate plans for all other players as an important byproduct). There is no reference to the planning player $k$ in this equilibrium problem --- in theory all players solve identical problems, though we introduce some mechanisms for recovery if this fails to be the case in sections \ref{sec:distribution_update} and \ref{sec:equilibrium_selection}.

\subsection{Motivating Opponent Information Tracking}

How is it that all players solve an identical game? In theory, a player does not know its opponents' observation histories, and it plays against opponents which do not know its own. To explain how Eq. \ref{eqn:succinct_tg} is formulated under this condition, we might imagine an optimization problem in which all players (as modeled by the planner) play in expectation over possible states and opponent observations, conditioned on their own observations. This leads to a version of $q$ which can be forward-updated via Bayesian filter. Since observations are private, each player is represented by an infinite number of optimization units in the equilibrium problem solved by player $k$ during planning, each with a unique possible observation history. We describe each optimization unit in this game as
\begin{equation}
    \left \{
        \argmin_{\theta^{(i, \hat{z}^{(i)})}}\;\;
        \int_{\substack{
            {x} \in \mathcal{X}^{1+\tfuture} \\
            {z} \in \mathcal{Z}^{1+\tfuture} \\
        }} 
        \hspace{-2.5mm}
        c^{(i)}(x)
        p(x, z | x_{\bar{t}}, \hat{z}^{(i)})
        q_{\bar{t}}(x_{\bar{t}}, \hat{z}^{(\neg i)} | \hat{z}^{(i)})
        \label{eqn:infinite_players_aig_tg}
    \right \}
\end{equation}
for \textit{every} possible observation history at planning time held by each player, ${\hat{z}^{(i)} \in (\mathcal{Z}^i)^\tpast}$. (This includes the planning player $k$ --- even though $\bar{z}^{(i)}$ is known by that player for $i=k$, the opponents modeled during planning must still treat it as unknown.)

Actualizing this setup would require agents to maintain the prior over both states and observation histories, $q_{\bar{t}}(x_{\bar{t}}, \hat{z}^{(\neg i)} | \hat{z}^{(i)})$, for every possible observation history $\hat{z}^{(i)}$ on which it might be conditioned. Of course, this is not possible, as there are infinitely many possible observation histories. Maintaining the fully joint probability over states and observation histories (which embeds $\hat{z}$) works around this. We use $q^{(i)}_{\bar{t}}(x_{\bar{t}}, \hat{z}^{(\neg i)} | \hat{z}^{(i)}) = q^{(i)}_{\bar{t}}(x_{\bar{t}}, \hat{z}) / p(\hat{z}^{(i)})$, and because player $i$ selects unique actions for all $\hat{z}^{(i)}$, any scaling factor of the form $f(\hat{z}^{(i)}) > 0$ in player $i$'s objective does not affect the policy parameters at equilibrium. As such, given $p(\bar{z}^{(i)}) > 0$, we may omit the marginal probability of $\hat{z}^{(i)}$ and consider $q_{\bar{t}}(x_{\bar{t}}, z_{[\bar{t}]})$, forming Eq. \ref{eqn:succinct_tg}.


\section{Approach}

Broadly, our approach maintains and updates $q^{(i)}_{\bar{t}}(x, z)$ using a particle representation, estimates the integral via Monte Carlo sampling, and solves the game with gradient play. Then, each player acts in accordance with model predictive game play, using their equilibrium policy with their true real-world observation history as input and advancing the true state of the world by only the first step of each plan. Players record a new real-world observation and the process is repeated.

The following sections describe this approach in detail.

\subsection{Equilibrium Computation}

We first turn to Nash equilibrium planning to solve the game presented in Eq. \ref{eqn:succinct_tg} given $q^{(i)}_{\bar{t}}(x_{\bar{t}}, z_{[\bar{t}]})$.

For each player, we evaluate the objective via ordinary Monte Carlo sampling, drawing $K_\textrm{batch}$ particles $(x_{\bar{t}}, z_{[\bar{t}]})$ from $q^{(i)}$ and then rolling out the game for $\tfuture$ steps on each particle to sample the remaining timesteps of $x$ and $z$, generating actions using each players' policy. Using automatic differentiation, the cost gradient with respect to the policy parameters $\nabla_{\theta^{(i)}} [c^{(i)}(x) \; p(x, z)]$ is calculated, and applied via gradient descent. Gradient steps are performed until convergence for all players. (All components of the game are differentiable per our Assumption 2). There are some conditions under which gradient play does not converge to a Nash equilibrium \cite{mazumdar2020gradient}; however, empirically, we find it converges reliably enough for online planning, as we will discuss in section \ref{sec:timing}.

As we will discuss, each player's $q^{(i)}_{\bar{t}}$ is implemented with a particle representation consisting of potential states $X$, potential finite observation histories $Z$, and weights $w$. With this representation, Algorithms \ref{alg:particle_cost} and \ref{alg:gradient_play} describe the equilibrium planning procedure in detail, with the former implementing the basic Monte Carlo objective estimate and the latter implementing gradient play over particles. $\texttt{opt}$ refers to any gradient-based optimization procedure on $\gamma$.

\begin{algorithm}
    \DontPrintSemicolon
    \SetKwInOut{Input}{Input}\SetKwInOut{Output}{Output}
    
    \Input{\,Particles ($X$, $Z$, $w$); parameters $\theta$; player $i$}
    \Output{\,Expected finite-horizon cost for $i$ across particles}

	\For{$k \in 1..K_\textrm{batch}$}{
        $x_{\bar{t}, k}, z_{[\bar{t}], k} \sim$ ($X$, $Z$) \, w.p. $w$ \;
        $c \leftarrow 0$ \;
        \While{$t < \bar{t} + \tfuture$\,}{
            $z_{t+1} \sim \mathcal{Z}$ \,w.p. $ O^{(i)}(\cdot | x_t)$ \;
            $a_{t_1} \leftarrow \pi_\theta(o_{[t+1]})$ \;
            $x_{t+1} \sim \mathcal{X}$ \,w.p. $ T^{(i)}(\cdot | x_{t}, a_{t+1})$  \;
            $c \leftarrow c + r^{(i)}(x_{t+1})$
        }
	}
   \Return{$c$}
   \caption{\texttt{obj} (expected cost over particles)\label{alg:particle_cost}}
\end{algorithm}

\begin{algorithm}
    \DontPrintSemicolon
    \SetKwInOut{Input}{Input}\SetKwInOut{Output}{Output}
	\SetKwFunction{Ecost}{\texttt{obj}}
	\SetKwFunction{Opt}{\texttt{opt}}
	\SetKwData{True}{\textbf{true}}
 
    \Input{\,Particles ($X$, $Z$, $w$); initialization $\theta$, tolerance $\epsilon$}
    \Output{\,Parameters in equilibrium}
    $c^{(i)}_0 \leftarrow \infty \;\; \forall i \in 1..N$ \;
    $j \leftarrow 1$

    \Repeat{$\delta^{(i)} < \epsilon \;\; \forall i \in 1..N$}{
        \For{$i \in 1..N$}{
            $\theta^{(i)} \leftarrow \Opt(\theta^{(i)}, \nabla_{\theta^{(i)}}$[\Ecost(($X$, $Z$, w), $\theta$, $i$)]) \;
            $c^{(i)}_j \leftarrow$ \Ecost(($X$, $Z$, w), $\theta$, $i$) \;
            $\delta^{(i)} \leftarrow c_j^{(i)} - c_{j-1}^{(i)} $
        }
        $j \leftarrow j + 1$
    }
   \Return{$\theta$}
   \caption{\texttt{calc\_eq} (gradient play over particles)\label{alg:gradient_play}}
\end{algorithm}

\subsection{Distribution Update}
\label{sec:distribution_update}

The planning component previously discussed requires knowledge of $q^{(i)}_{\bar{t}}$, the distribution over game histories at the current time according to player $i$, necessitating a suitable update rule. Intuitively, $q^{(i)}_{\bar{t}}$ is the joint distribution of state/observation histories at time $\bar{t}$ assuming that all players have behaved rationally up to time $\bar{t}$. Since agents find equilibrium policies at each time step, $q^{(i)}_{\bar{t}}$ can be updated in an open-loop manner using only $\pi$, completely ignoring real-world observations. This is possible because $\pi$ maps all possible observation histories to an action, not just $\bar{z}_{[\bar{t}]}$. By maintaining $q^{(i)}_{\bar{t}}$ as the joint distribution over histories of observations and states, players need not track a belief hierarchy over each others' beliefs: this joint distribution is the same for all players, and players' actions are determined entirely by the observations they receive, to which they respond optimally.

To that end, we approximate $q^{(i)}_{\bar{t}}$ with $K_\textrm{all} \gg K_\textrm{batch}$ particles $(X, Z, w)$, with $X_k = x_{\bar{t}, k}$ and $Z_k = z_{[\bar{t}], k}$ for $k \in 1.. K_\textrm{all}$, each particle with weight $w_{k}$. After planning, each particle is stepped forward with the equilibrium policies and receives a sampled observation for each player according to its joint state. Particle weights are not affected, nor is there any conditioning on $\bar{z}_{\bar{t}}$.

We use this unconditioned approach because any approach conditioning $q^{(i)}_{\bar{t}}$ with $\bar{z}_{\bar{t}}$ causes an asymmetry between agents, which results in planning against opponents that are not representative of the real world (and therefore a suboptimal plan). However, if any player diverges from the equilibrium, or players find different equilibria, particles are updated with unrepresentative equilibrium policies and $q^{(i)}_{\bar{t}}$ may no longer match the true state. Furthermore, with limited particles, it is possible that the planning player's true observations $\bar{z}^{(i)}$ may not be represented jointly with the true state in any particle, causing the policy for $\bar{z}^{(i)}$ to optimize poorly during gradient play. As such, there is a fundamental trade-off between usage of true observations to model the world state efficiently, and correctness in the corresponding model of the opponent.

Typically, a small amount of closed-loop behavior is desirable, even at the cost of inaccurate opponent representations. To accomplish this, we condition a small proportion of particles $\gamma$ with true observations, reweighting them accordingly. $\gamma$ can be interpreted as modulating the closed-loop / opponent-modeling trade-off.

Algorithm \ref{alg:particle_update} implements this distribution update.

\begin{algorithm}
    \DontPrintSemicolon
    \SetKwInOut{Input}{Input}\SetKwInOut{Output}{Output}
    \SetKwBlock{WithProb}{with probability}{}
    
    \Input{\,Particles ($X$, $Z$, $w$), true observation $\bar{z}_{\bar{t}}^{(i)}$, policy parameters $\theta$, probability $\gamma$}
    \Output{\,Updated particles ($X'$, $Z'$, $w'$)}


    \For{$k \in 1..K_\textrm{all}$}{
        $z'_{\bar{t}} \sim \mathcal{Z}$ \,w.p. $O(\cdot | X_k)$ \;
        $w'_k \leftarrow w_k$ \;
        \WithProb($\gamma$){
            $z'_{\bar{t}} \leftarrow [\bar{z}_{\bar{t}}^{(i)}, z_{\bar{t}}'^{(\neg i)}]$ \;
            $w'_k \leftarrow w_k \cdot O(\bar{z}_{\bar{t}} | X_k)$ \;
        }
        $Z_k' \leftarrow [Z_k..., z'_{\bar{t}}]_{[\bar{t}]}$ \;
        $X_k' \sim \mathcal{X}$ \,w.p. $T(\cdot | X_k, \pi(z'; \theta))$
    }
    \Return{($X'$, $Z'$, $w'$)}
    
    \caption{\texttt{update\_particles} (update rule for $q^{(i)}$)\label{alg:particle_update}}
\end{algorithm}

\vspace{-0.5cm}

\subsection{Equilibrium Selection}
\label{sec:equilibrium_selection}

Our method can be deployed in a ``shared brain'' setting, in which case it calculates an equilibrium joint policy once across all players, or in a ``separate brain'' setting, in which case all players do so independently. In the former case, it is assumed that $q^{(i)}_{\bar{t}} = q^{(j)}_{\bar{t}}$ for all $i, j \in 1..N$. However, in the latter, more realistic case where agents plan individually, this is not generally true. Agents may find different local equilibrium joint policies, potentially modeling opponent behavior incorrectly and causing $q^{(i)}_{\bar{t}}$ and $q^{(j)}_{\bar{t}}$ to diverge for $j \neq i$. 

\begin{figure}[H]
    \centering
    \includegraphics[width=\linewidth,trim={0 0.3cm 0 0.3cm}]{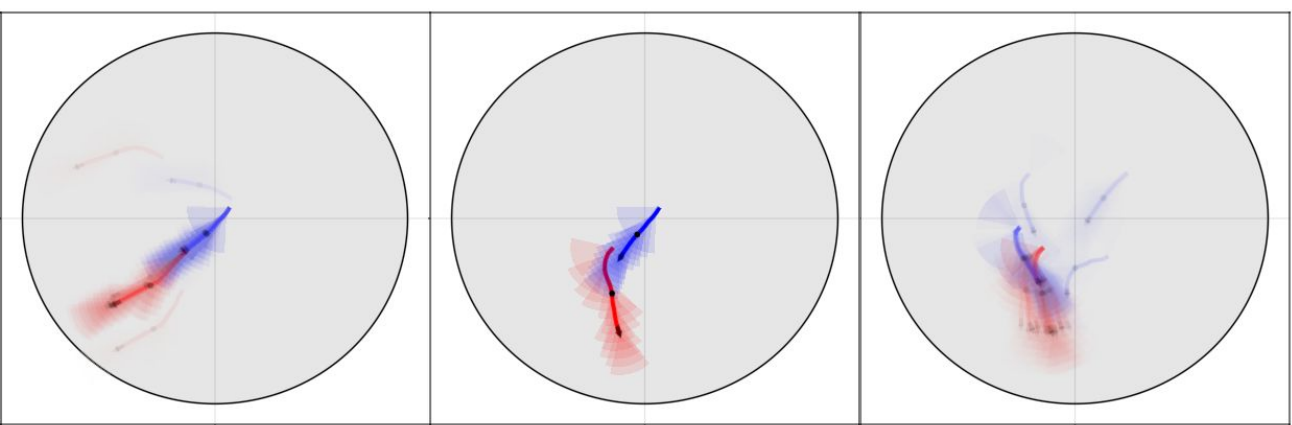}
    \caption{Visualization of an equilibrium selection failure in a pursuit-evasion game. Opacity mapped to particle weight. \\(Left): Pursuer $q^{(i)}$.\;\;(Middle): True state.\;\;(Right): Evader $q^{(i)}$. }
    \label{fig:dists_diverge}
\end{figure}

Figure \ref{fig:dists_diverge} visualizes this failure case in a pursuit-evasion scenario (for more detail on the pursuit-evasion game used here, see Section \ref{sec:experiments}). The left and right plots visualize the pursuer (blue) and evader (red) distributions respectively. In this case, the pursuer finds one deterministic-strategy equilibrium in which the evader turns right at the boundary, and the evader finds a different equilibrium in which it turns left. The discrepancy primarily harms the pursuer, causing it to pursue an inaccurate model of the evader.

Fortunately, as long as one or more particles behave consistently with an opponent's true policy and $\gamma > 0$, the correct $D^{(i)}$ can generally be recovered as conditional updates deweight particles that are inconsistent with observations of the opponent's true state evolution in favor of those that are. To promote the existence of particles following different possible equilibria selections, an agent can complete the equilibrium planning step $N_\textrm{eq}$ times. Each of these candidate policies is applied to a subset of particles during the distribution update step, and the first one is arbitrarily chosen to provide the real-world action.

Alg. \ref{alg:ourmethod} fully describes our proposed method in the separate-brain setting. In Alg. \ref{alg:ourmethod}, \texttt{partition} creates $N_\textrm{eq}$ roughly-equal index sets of the particles, and \texttt{act} performs an action in the real world; the remaining functions are as given in Alg.s \ref{alg:gradient_play} and \ref{alg:particle_update}. Note that Alg. \ref{alg:ourmethod} is not always guaranteed to converge for all POSGs --- in particular, those which do not have pure strategy equilibria (violating Assumption 3). However, we find its empirical performance compelling, as we demonstrate over the following two sections.

\begin{algorithm}
    \DontPrintSemicolon
    \SetKwInOut{Input}{Input}\SetKwInOut{Output}{Output}
    \SetKwBlock{WithProb}{with probability}{}
    \SetKwComment{Comment}{/* }{ */}
	\SetKwFunction{Act}{\texttt{act}}
	\SetKwFunction{Update}{\texttt{update\_particles}}
	\SetKwFunction{Partition}{\texttt{partition}}
	\SetKwFunction{Solve}{\texttt{calc\_eq}}
    \For{$k \in 1..K_\textrm{all}$}{ 
        $X_k \sim X_0$ \;
        $Z_k \leftarrow \vec{0}_{\tpast \times |\mathcal{O}|}$ \;
    }
    $\mathcal{P} \leftarrow \Partition(1..K_{all}, N_\textrm{eq})$ 
    \;
    $\bar{z}^{(i)} \leftarrow \vec{0}_{\tpast \times |\mathcal{O}^{(i)}|}$ \;
    \;

    \For{$\bar{t} \in 1..\infty$}{
        \For{$p \in 1..N_\textrm{eq}$}{
            $\theta_p \leftarrow \Solve((X, Z, w), \theta_p)$ \; 
        } 
        
        $\bar{z}_{\bar{t}}^{(i)} \leftarrow \Act(\pi^{(i)}_{\theta^{(i)}_1}(\bar{z}_{[\bar{t}]}))$ \;

        \For{$p \in 1..N_\textrm{eq}$}{
        $P \leftarrow \mathcal{P}_p$ \;
            $(X_P, Z_P, w_P) \leftarrow \,\,\,\Update((X_P, Z_P, w_P), \bar{z}^{(i)}_{\bar{t}}, \theta_p$) \; 
        }
    }
    \caption{Active information planning}
    \label{alg:ourmethod}
\end{algorithm}

\subsection{Implementation}
We implemented this approach in the Julia programming language \cite{bezanson2017julia}. Policies are represented with feedforward neural networks and optimized with the Adam optimizer. We used Flux.jl \cite{innes2018flux} to implement these networks and perform automatic differentiation.



\section{Experimental Setup}
\label{sec:experiments}

To test our method, we considered three variants of a UAV pursuit-evasion scenario as well as a warehouse-pickup scenario, all in continuous space. We describe these scenarios in this section.

\begin{figure*}[h]
    \centering
    \includegraphics[width=0.9\linewidth]{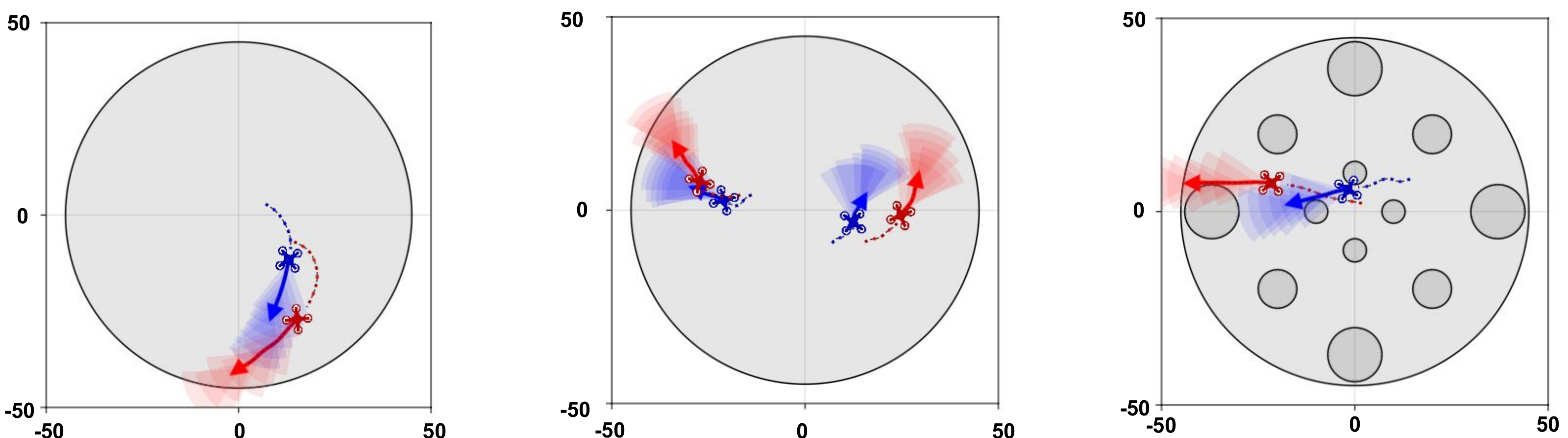}
    \caption{Trajectories in the three simulated UAV tag environments. Pursuer is blue; evader is red. Plans are solid lines and histories are dotted. \\(Left) Field of view tag. \;\;(Center) Tag chain. \;\;(Right) Hide and seek.}
    \label{fig:tag_games}
\end{figure*}

\subsection{Pursuit-Evasion (UAV Tag)}
Pursuit-evasion is a self-explanatory scenario in two dimensions comparable to the childhood game of ``tag.'' We simulated two UAVs equipped with fixed field-of-view cameras, notated ``pursuer'' and ``evader''.

\textbf{Control.} Agents control their own velocity $x_\textrm{vel}$, and their position $x_\textrm{pos}$ is simulated through double-integrator dynamics for $\tfuture = \tpast = 7$ timesteps. The evader has a slightly greater maximum velocity than the pursuer.

\textbf{Costs.} The pursuer (arbitrarily indexed $i=1$) seeks to minimize $||x_\textrm{pos}^{(1)} - x_\textrm{pos}^{(2)}||_2$ while the evader seeks to maximize it at every timestep. Agents are independently penalized for departures from a circular area; otherwise, this is a zero-sum game. 

\textbf{Observations.} Both players receive perfect observations of their own state. Observations of opponent position are limited to a field of view in the UAV's direction of motion. Specifically, observations of the opponent are drawn from a trimmed multivariate Gaussian around the opponent location with $\mu = x_\textrm{pos}^{(\neg i)}$ and

\begin{equation}
    \sigma^2 = \sigma^2_\textrm{base} + \begin{cases}
         C_\textrm{scale} \cdot (|\theta| - \frac{f}{2}) & |\theta| \geq \frac{f}{2} \\
        0 & \textrm{otherwise}
    \end{cases}
\end{equation}

where $\theta$ is the bearing between the agent's heading (taken from their velocity) and their opponent, and $f$ is the agent's field of view, both in radians. Observations are trimmed to be within the play area. Parameters $\sigma^2_\textrm{base}$ and $C_\textrm{scale}$ adjust the minimum variance and the variance per radian outside the field of view, respectively.

\textbf{Initialization.} Initial states are normally distributed around the origin. There is no process noise; uncertainty is a result of initial uncertainty or observation noise (as a simplification, rather than a limitation of the method).

\textbf{Variant: Tag chain.} In this variant, there are any (even) number of players: $\frac{N}{2}$ pursuers labeled $P_1..P_{\frac{N}{2}}$ pursue evaders $E_1..E_{\frac{N}{2}}$. Pursuer $P_i$ pursues evader $E_i$, but evader $E_i$ evades pursuer $P_{i+1}$ (and evader $E_{\frac{N}{2}}$ evades pursuer $P_1$). This results in a cycle of partial pursuit-evasion games. Agents receive observations of all other agents in their field of view. We consider this variant to demonstrate extensibility to $N$-player, general-sum scenarios. We use $N=4$, and for simplicity, we consider chain tag in the ``shared brain'' setting only, assuming $q^{(i)}_{\bar{t}} = q^{(j)}_{\bar{t}}$.

\textbf{Variant: Hide and seek.} In this variant, the simulated play area includes circular obstacles that block UAVs' fields of view. Agents receive high-variance observations for obstructed opponents in the same way as those outside the field of view. A penalty is applied for colliding with these obstacles in the same manner as the exterior boundary. We consider this variant to showcase a complex environment which may incentivize nuanced hiding and detection behavior.

\subsection{Warehouse Pickup}
We also consider one non-pursuit-evasion example. Imagine a warehouse which uses an autonomous robot (``P1'') to load goods. P1 simply proceeds greedily towards the nearest task location. At a later time, an additional robot (``P2''), created by a different manufacturer, is introduced. Unlike P1, P2 is designed for multi-robot environments and does not proceed greedily towards task location, instead reasoning about other agents in a decentralized manner. P2 is connected to a simple station which broadcasts the locations of both robots.

We seek a plan for P2 that works around P1's inability to cooperate. However, there is an additional wrinkle --- locations broadcast by the station may be noisy. This inaccuracy is modeled as the sum of two components: inaccuracy caused by potentially stale data about a distant P1, and inaccuracy caused by channel noise between P2 and the station. We use P1's and P2's distance from the station, respectively, as proxies for the amount of noise incurred by these effects.

\textbf{Control.} The warehouse is modeled by the $[0, 1]^2$ space. Control is the same as in pursuit-evasion examples. P2 has a higher maximum velocity than P1.

\textbf{Costs.} Both players are penalized by their distance from obstacles, with additive penalty $-\exp(-\beta ||x_\textrm{pos}^{(i)} - \tau_j||_2^2)$ for each target $j$. P2 is additionally penalized for proximity to P1, with additive penalty $-\alpha \exp(-\beta ||x_\textrm{pos}^{(2)} - x_\textrm{pos}^{(1)}||_2^2)$. $\alpha$ and $\beta$ control the weighting between P2's objectives and the scale used to determine proximity; we use $\alpha=4.0$ and $\beta=20$.

\textbf{Observations.} P1 only observes its own position perfectly. P2 observes its own position perfectly and P1's position imperfectly. Its observations of P1 are exact when both players are at the station (notated $s := [0.5, 1.0]$) and receive additive noise $\sim \mathcal{N}(0, \sigma^2)$ where $\sigma = \eta_1||x^{(1)}_\textrm{pos} - s||_2 + \eta_2||x^{(2)}_\textrm{pos} - s||_2$. $\eta$ controls the strength of each noise component; here, we use $\eta_1=\eta_2=4$.

\textbf{Initialization.} Both robots are randomly initialized anywhere in the warehouse. We use two task locations, which are also random.

\begin{figure}[h]
    \centering
    \includegraphics[width=0.8\linewidth]{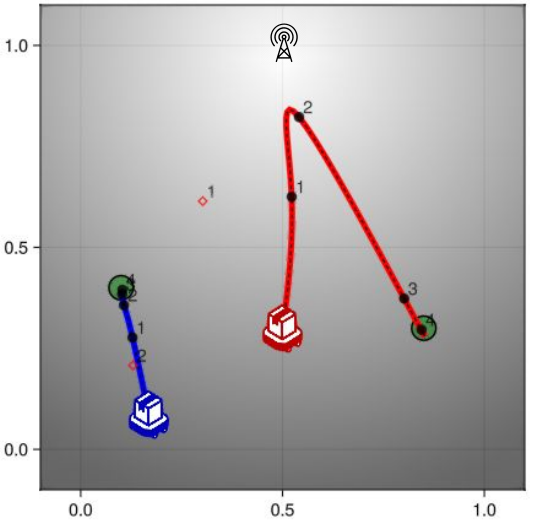}
    \caption{Active information gathering behavior in a 4-step warehouse pickup environment. Background gradient indicates observation accuracy. Diamonds indicate P2's observations (observations 3 and 4 are out of bounds). Green circles indicate task locations.}
    \label{fig:warehouse}
\end{figure}

Fig. \ref{fig:warehouse} visualizes correct behavior in the warehouse pickup scenario (as generated by Alg. \ref{alg:gradient_play}). P2 (red) first travels to the station to localize P1, then selects the task location further from P1 to prevent a collision.

\section{Results}

\subsection{Improvement from Active Information Gathering}
\label{sec:evidence_of_aig}

As a baseline, we used an passive-gathering version of our solver which implements a variant of Eq. \ref{eqn:succinct_tg} where $\pi$ maps only from past observations. We ran each of the four possible configurations --- passive or active solver for each player --- on twenty independent trials for each scenario (seventy for the warehouse example). Each lasted twenty timesteps. We used $\tfuture=\tpast=6$, $\gamma=0.1$, $K_\textrm{all}=1000$, and $K_\textrm{batch}=10$, except where noted.  

For simplicity of analysis, we assume equilibrium selection succeeds and players find the same equilibrium --- we explicitly consider equilibrium selection in the next section. We terminate Alg. \ref{alg:gradient_play} after 100 gradient iterations (rather than allowing convergence) for a more level comparison between active and passive information gathering, as the active version typically takes more iterations to converge. 

The competitive behavior of the agents for each active/passive configuration and each scenario is quantified in Table \ref{tab:concise_results}, listing the average costs (in distance units) achieved by each player over all trials with standard error. We do not include boundary penalties in the reported cost. Both pursuer/P1 (blue) and evader/P2 (red) costs are shown in each cell. We note the following scenario-specific details: (1) \textsc{Tag} and \textsc{Hide\&Seek} are effectively zero-sum, thus the apparent repetition. (2) In \textsc{TagChain}, mean costs for each team (evaders and pursuers) are used rather than individual costs. (3) In the \textsc{Warehouse} example, P1 has no imperfect observations and therefore there is no distinction between active and passive modes.

\textbf{In all cases except one, our method for active information gathering elicits a reduction in cost}. In Table \ref{tab:concise_results}, blue (pursuer) costs are lower in the left column, which corresponds to that agent using active information gathering, and likewise red (evader/P2) costs are lower in the second row of each scenario. Figures \ref{fig:passive_vs_active_gathering} and \ref{fig:warehouse} show the desirable, active plans in \textsc{Tag} and \textsc{Warehouse}, where in both cases an information-gathering trajectory clearly outperforms a greedy plan. Several of these differences (particularly in \textsc{Tag} and \textsc{Warehouse}) are statistically significant with $p < 0.05$. 

The single exception to improvement is the evader in \textsc{Hide\&Seek}, which exhibits a cost increase under our method. We reason that optimal evader strategy in this game is too complex to be captured with only $\tpast$ past observations, and therefore modeling information in any given 6-step horizon is a poor strategic decision which unnecessarily complicates the optimization procedure.

\begin{table}[h]
\begin{center}
\begin{tabular}{@{}llll@{}}
                                              &                                                     & \multicolumn{1}{c}{\textbf{\begin{tabular}[c]{@{}c@{}}Passive \\ \blue{pursuer}\end{tabular}}}                    & \multicolumn{1}{c}{\textbf{\begin{tabular}[c]{@{}c@{}}Active\\ \blue{pursuer}\end{tabular}}}                      \\ \cmidrule(l){3-4} 
\multirow{2}{*}{\textsc{\rotate{Tag}}}        & \multicolumn{1}{l|}{\textbf{Passive \red{evader}}}  & \begin{tabular}[c]{@{}l@{}}\blue{12.58 } \lblue{ ± 1.44}\\ \red{-12.58 } \lred{ ± 1.44}\end{tabular}                 & \begin{tabular}[c]{@{}l@{}}\blue{9.97 } \lblue{ ± 1.19}\\ \red{-9.97 } \lred{ ± 1.19}\end{tabular}                   \\
                                              & \multicolumn{1}{l|}{\textbf{Active \red{evader}}}   & \begin{tabular}[c]{@{}l@{}}\blue{13.97 } \lblue{ ± 1.92}\\ \red{-13.97 } \lred{ ± 1.92} \vspace{0.05cm}\end{tabular} & \begin{tabular}[c]{@{}l@{}}\blue{12.98 } \lblue{ ± 1.18}\\ \red{-12.98 } \lred{ ± 1.18} \vspace{0.05cm}\end{tabular} \\
                                              & \vspace{0.2cm}                                      &                                                                                                                      &                                                                                                                      \\
\multirow{2}{*}{\textsc{\rotate{TagChain}}}   & \multicolumn{1}{l|}{\textbf{Passive \red{evaders}}} & \begin{tabular}[c]{@{}l@{}}\blue{17.68 } \lblue{ ± 1.39}\\ \red{-25.91 } \lred{ ± 1.54}\end{tabular}                 & \begin{tabular}[c]{@{}l@{}}\blue{16.79 } \lblue{ ± 1.43}\\ \red{-25.12 } \lred{ ± 1.57}\end{tabular}                 \\
                                              & \multicolumn{1}{l|}{\textbf{Active \red{evaders}}}  & \begin{tabular}[c]{@{}l@{}}\blue{15.79 } \lblue{ ± 1.23}\\ \red{-27.48 } \lred{ ± 1.39} \vspace{0.05cm}\end{tabular} & \begin{tabular}[c]{@{}l@{}}\blue{15.71 } \lblue{ ± 1.26}\\ \red{-27.29 } \lred{ ± 1.41} \vspace{0.05cm}\end{tabular} \\
                                              & \vspace{0.2cm}                                      &                                                                                                                      &                                                                                                                      \\
\multirow{2}{*}{\textsc{\rotate{Hide\&Seek}}} & \multicolumn{1}{l|}{\textbf{Passive \red{evader}}}  & \begin{tabular}[c]{@{}l@{}}\blue{19.62 } \lblue{ ± 1.33}\\ \red{-19.62 } \lred{ ± 1.33}\end{tabular}                 & \begin{tabular}[c]{@{}l@{}}\blue{17.93 } \lblue{ ± 1.34}\\ \red{-17.93 } \lred{ ± 1.34}\end{tabular}                 \\
                                              & \multicolumn{1}{l|}{\textbf{Active \red{evader}}}   & \begin{tabular}[c]{@{}l@{}}\blue{18.91 } \lblue{ ± 1.25}\\ \red{-18.91 } \lred{ ± 1.25} \vspace{0.05cm}\end{tabular} & \begin{tabular}[c]{@{}l@{}}\blue{16.75 } \lblue{ ± 1.34}\\ \red{-16.75 } \lred{ ± 1.34} \vspace{0.05cm}\end{tabular} \\
                                              & \vspace{0.2cm}                                      &                                                                                                                      &                                                                                                                      \\
                                              &                                                     & \multicolumn{1}{c}{\textbf{Passive \blue{P1}}}                                                                       & \multicolumn{1}{c}{\textbf{Active \blue{P1}}}                                                                        \\ \cmidrule(l){3-4} 
\multirow{2}{*}{\textsc{\rotate{Warehouse}}}  & \multicolumn{1}{l|}{\textbf{Passive \red{P2}}}      & \begin{tabular}[c]{@{}l@{}}\blue{-5.12 } \lblue{ ± 0.26}\\ \red{-0.02 } \lred{ ± 0.34}\vspace{0.05cm}\end{tabular}   & \multicolumn{1}{c}{n/a}                                                                                              \\
                                              & \multicolumn{1}{l|}{\textbf{Active \red{P2}}}       & \begin{tabular}[c]{@{}l@{}}\blue{-5.21 } \lblue{ ± 0.26}\\ \red{-1.76 } \lred{ ± 0.54}\end{tabular}                  & \multicolumn{1}{c}{n/a}                                                                                             
\end{tabular}
\caption{Costs in passive/active configurations, per scenario}
\label{tab:concise_results}
\end{center}
\end{table}

\subsection{Timing and Convergence}
\label{sec:timing}
While this method is not yet real-time, we do report some preliminary timing results. 

Table \ref{tab:timing} gives the average wall time \textit{per gradient step}, per player, for each scenario. We average over 20 trials, otherwise using the same setup as in the previous experiments. (Note that the particle distribution updates take a negligible amount of time in comparison). Based on these results, we believe real-time execution is within reach, particularly as code optimization, GPU support, better handling of matrix sparsity, and improved parallelization all remain as potential improvements. 

\begin{table}[h]
\begin{center}
\begin{tabular}{ll}
\textbf{Scenario}   & \textbf{\begin{tabular}[c]{@{}l@{}}Time (seconds) \vspace{0.1cm}\end{tabular}} \\
\textsc{Tag}        & 0.085 ± 0.036                                                                                          \\
\textsc{TagChain}   & 0.222 ± 0.078                                                                                          \\
\textsc{Hide\&Seek} & 0.117 ± 0.030                                                                                          \\
\textsc{Warehouse}  & 0.062 ± 0.039

\end{tabular}
\caption{Optimization times per gradient step, per player}
\label{tab:timing}
\end{center}
\end{table}

\begin{figure}[h]
    \centering
    \includegraphics[width=0.9\linewidth]{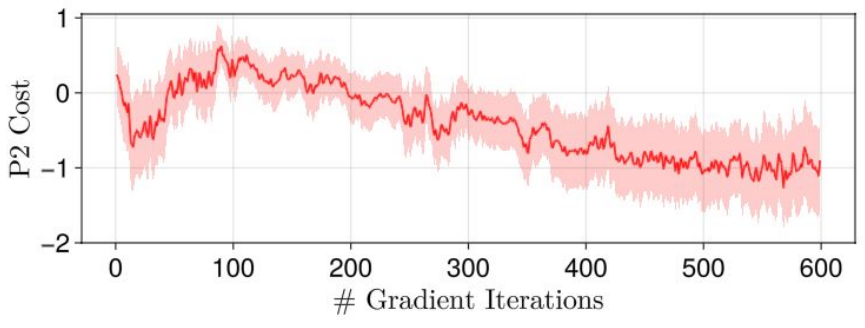}
    \caption{P2's cost as the first step of \textsc{Warehouse} is solved. Ribbon indicates standard error.}
    \label{fig:convergence}
\end{figure}

As noted previously, Alg. \ref{alg:ourmethod} is not guaranteed to converge in all scenarios --- for instance, some configurations of pursuit-evasion have mixed strategy solutions which cannot be found using pure gradient play. However, empirically, we note that the algorithm converges well when pure strategy solutions are available. Figure \ref{fig:convergence} shows the cost convergence when optimizing the first step of \textsc{Warehouse} over 600 gradient iterations, taken across ten trials, against a randomly placed P1. While there is still inherent noise due to the method of stochastic gradient play, the method largely converges, and as policy parameters from the previous timestep are reused as initial parameters, subsequent steps benefit from past optimization (motivating our choice of 100 gradient steps in our experiments).

We also examined empirical scaling with respect to the method parameters. Table \ref{tab:scaling_tfuture} shows scaling behavior w.r.t $T_\textrm{future}$ over ten trials. The relation between $T_\textrm{future}$ and time-to-converge is theoretically exponential: the number of possible world configurations, and ergo particles that must be processed, explodes with time. However, information gathering early in the plan limits potential future configurations (and reduces the future cost). (On the other hand, we found little relation between $T_\textrm{past}$ and time-to-converge, which we believe to be specific to the relatively atemporal games we analyzed.)

\begin{table}[H]
\centering
\begin{tabular}{lll}
\textbf{$T_\textrm{future}$} & \textbf{Cost} & \textbf{Wall Time (s)} \\
1                                            & 0.604 ± 0.21  & 4.411 ± 0.29  \\
2                                            & 0.338 ± 0.11  & 7.907 ± 1.35  \\
3                                            & 0.253 ± 0.09  & 7.254 ± 1.03  \\
4                                            & 0.023 ± 0.10  & 11.281 ± 1.33 \\
5                                            & -0.075 ± 0.11 & 12.009 ± 1.85 \\
6                                            & -0.197 ± 0.10 & 16.415 ± 2.91 \\
7                                            & -0.162 ± 0.13 & 19.509 ± 2.67
\end{tabular}
\caption{\vspace{-4mm}}
\label{tab:scaling_tfuture}
\end{table}

Finally, we report scaling with respect to the number of particles per batch $K_\textrm{batch}$. ($K_\textrm{all}$ has little effect on convergence, as gradient play, the most expensive component, only occurs over the batch). Insufficient $K_\textrm{batch}$ results in fast convergence to an information-unaware minimum, as indicated in Table \ref{tab:scaling_kbatch}. 

\begin{table}[H]
\centering
\begin{tabular}{lll}
\textbf{$K_\textrm{batch}$} & \textbf{P2 Cost} & \textbf{Wall Time (s)} \\
2                           & 0.326 ± 0.02     & 4.86 ± 0.29            \\
6                           & 0.308 ± 0.02     & 5.692 ± 1.11           \\
15                          & 0.227 ± 0.03     & 7.199 ± 1.18           \\
39                          & 0.206 ± 0.05     & 7.673 ± 0.84           \\
100                         & 0.067 ± 0.04     & 11.791 ± 1.5           \\
251                         & 0.052 ± 0.04     & 14.462 ± 1.79          \\
630                         & 0.001 ± 0.07     & 32.497 ± 4.65         
\end{tabular}
\caption{\vspace{-4mm}}
\label{tab:scaling_kbatch}
\end{table}

\subsection{Equilibrium Agreement}
\balance
In general, POSGs can admit arbitrarily many equilibria in deterministic strategies, and determining which equilibrium an opponent is playing is a fundamental difficulty of agent modeling. Section \ref{sec:equilibrium_selection} details our approach to minimizing this issue. In this section we analyze how well this approach performs.

\begin{figure}[h]
    \centering
    \includegraphics[width=\linewidth]{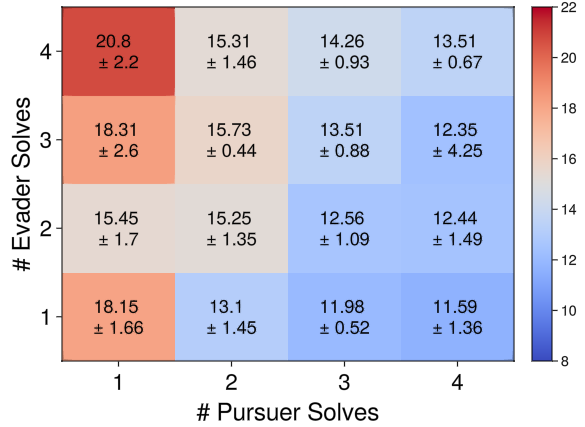}
    \caption{Distance between players in field of view tag for different configurations of $N_\textrm{eq}$ Red configurations favor the evader; blue favor the pursuer.}
    \label{fig:nsolves_heatmap}
\end{figure}

We tested ordinary field of view tag with different settings of $N_\textrm{eq}$ for each player, playing the game for 100 steps but otherwise keeping the same setup as Section \ref{sec:evidence_of_aig}. Figure \ref{fig:nsolves_heatmap} shows distance between the pursuer and evader for each configuration of $N_\textrm{eq}$ (blue configurations favor the pursuer; red favor the evader). Performance improves for both players with more equilibrium solves. Games with insufficient $N_\textrm{eq}$ favor the evader, possibly because pursuer equilibrium policies (which converge on the evader) are more similar than evader equilibrium policies (which diverge).

\begin{figure}[h]
    \centering
    \includegraphics[width=0.9\linewidth,trim={0 1.2cm 0 1cm}]{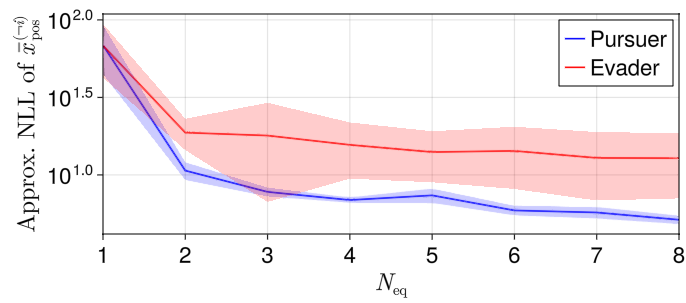}
    \caption{Negative log likelihood of opponent position for settings of $N_\textrm{eq}$. Logarithmic scale. Ribbons indicate standard error.}
    \label{fig:neq_surprisal}
\end{figure}

To confirm that this improvement is actually caused by better opponent modeling, we tracked the approximate expected negative log likelihood (surprisal) of the true opponent position for various settings of $N_\textrm{eq}$ for each agent, using a Gaussian approximation of $q^{(i)}_{\bar{t}}(\bar{x}^{(\neg i)}_{\bar{t}})$ and taking $N_\textrm{eq} = 4$ for the opponent. Higher values indicate the opponent true position is not captured by $q_{\bar{t}}^{(i)}$. Figure \ref{fig:neq_surprisal} shows the results, indicating that a greater number of solves decreases uncertainty about the opponent position. The pursuer tends to receive more informative observations, improving its estimates independently of $N_\textrm{eq}$; this causes the discrepancy between pursuer and evader seen in Figure \ref{fig:neq_surprisal}. Note that trials for each setting of $N_\textrm{eq}$ are seeded by sample number.

\section{Conclusion}

\subsection{Summary}
In this work we discussed rational information gathering in the context of model predictive game planning. We formulated the active planning problem in a finite-history/horizon setting, and described a basic method for incorporating potential future observations into the planning procedure using gradient play and a number of approximations and additions to make the method suitable for independent planning agents. Using that method, we showed that active information gathering is indeed competitively preferable for players in variants of a pursuit-evasion game, and provided additional analysis on the parameters of the solution method.

\subsection{Limitations} 
As noted, we implemented our approach with no attention given to fast planning. In reality, code and hardware optimizations make this prospect more likely, making this approach ``online'' in both the analytical and practical sense.

More fundamentally, we have assumed a finite history of observations is suitably informative, and as we discussed regarding the ``hide-and-seek'' example, performance suffers when this is not the case. There are many potential approaches to handling the curse of history, and applying any of them in a way that avoids the need for recursively hierarchical beliefs is a promising research direction. 

The method is also subject to all assumptions made in Section \ref{sec:posgs_framework}. In particular, nondeterministic strategies can be permitted with a minor formulation change, and the ability to deliberately, directly introduce uncertainty into the state is quite consequential in settings permitting information gathering (and also potentially permits theoretical convergence guarantees). Combining imperfect information methods with mixed-strategy approaches like that in \cite{peters2022learning} may be a source of further improvement in competitive play.


\bibliographystyle{plainnat}
\bibliography{bibliography}

\end{document}